\documentclass[lettersize,journal]{IEEEtran}
\usepackage{amsmath,amsfonts}
\usepackage{multirow}
\usepackage{array}
\usepackage{enumerate}
\usepackage[utf8]{inputenc}
\usepackage[T1]{fontenc}
\usepackage{amsfonts}
\usepackage{amssymb}
\usepackage{xcolor}
\usepackage{enumitem}
\usepackage{algorithm}  
\usepackage{algpseudocode}  
\usepackage{amsmath}  
\usepackage{breqn}
\usepackage{textcomp}
\usepackage{stfloats}
\usepackage{url}
\usepackage{verbatim}
\usepackage{graphicx}
\usepackage{cite}
\usepackage{enumitem}
\usepackage[caption=false]{subfig}
\usepackage{hyperref}
\usepackage{tipx}
\hypersetup{
    colorlinks=true,
    linkcolor=blue,
    filecolor=magenta,      
    urlcolor=cyan,
    }
\begin{document}
\setlength{\abovedisplayskip}{1pt} 
\setlength{\belowdisplayskip}{1pt} 

\title{\LARGE \bf  \textcolor{black}{Multimodal Audio-based Disease Prediction with \\ Transformer-based Hierarchical Fusion Network}
 }

\author{Jinjin Cai$\dag^1$, Ruiqi Wang$\dag^1$, Dezhong Zhao$^{1,2}$, Ziqin Yuan$^{1,3}$, Victoria McKenna$^4$, Aaron Friedman$^5$, \\Rachel Foot$^5$, Susan Storey$^6$, Ryan Boente$^7$, Sudip Vhaduri$^1$, and Byung-Cheol Min$^1$ 

\thanks{$\dag$ Equal contribution.}
\thanks{$^1$Department of Computer and Information Technology, Purdue University, West Lafayette, IN, USA. {\tt\small{[cai379, wang5357, svhaduri, minb]@purdue.edu}.}}
\thanks{$^{2}$College of Mechanical and Electrical Engineering, Beijing University of Chemical Technology, Beijing, China. \tt\small{DZ\_Zhao@buct.edu.cn}.}
\thanks{$^{3}$Tandon School of Engineering, New York University, Brooklyn, NY, USA. \tt\small{zy2232@nyu.edu}.}

\thanks{$^4$Department of Communication Sciences and Disorders, Biomedical Engineering, Otolaryngology-Head \& Neck Surgery, University of Cincinnati, Cincinnati, OH 45267. {\tt\small{mckennvs@ucmail.uc.edu}.}}

\thanks{$^5$College of Medicine, University of Cincinnati, Cincinnati, OH 45267. {\tt\small{[friedma5, footrl]@ucmail.uc.edu}.}}

\thanks{$^6$School of Nursing, Indiana University, Indianapolis, IN 46202. {\tt\small{sustorey@iu.edu}.}}

\thanks{$^7$School of Medicine, Indiana University, Indianapolis, IN 46202. {\tt\small{rboente@iu.edu}.}}

}

\markboth{}%
{Shell \MakeLowercase{\textit{et al.}}: A Sample Article Using IEEEtran.cls for IEEE Journals}

\IEEEpubid{}

\maketitle

\begin{abstract}
Audio-based disease prediction is emerging as a promising supplement to traditional medical diagnosis methods, facilitating early, convenient, and non-invasive disease detection and prevention. Multimodal fusion, which integrates features from various domains within or across bio-acoustic modalities, has proven effective in enhancing diagnostic performance. However, most existing methods in the field employ unilateral fusion strategies that focus solely on either intra-modal or inter-modal fusion. This approach limits the full exploitation of the complementary nature of diverse acoustic feature domains and bio-acoustic modalities. Additionally, the inadequate and isolated exploration of latent dependencies within modality-specific and modality-shared spaces curtails their capacity to manage the inherent heterogeneity in multimodal data. To fill these gaps, we propose a transformer-based hierarchical fusion network designed for general multimodal audio-based disease prediction. Specifically, we seamlessly integrate intra-modal and inter-modal fusion in a hierarchical manner and proficiently encode the necessary intra-modal and inter-modal complementary correlations, respectively. Comprehensive experiments demonstrate that our model achieves state-of-the-art performance in predicting three diseases: COVID-19, Parkinson's disease, and pathological dysarthria, showcasing its promising potential in a broad context of audio-based disease prediction tasks. Additionally, extensive ablation studies and qualitative analyses highlight the significant benefits of each main component within our model.

\end{abstract}

\begin{IEEEkeywords}
Hierarchical Data Fusion, Multimodal Deep Learning, Audio-based Disease Prediction, Speech Analysis, Parkinson’s Disease, COVID-19 Diagnostics.
\end{IEEEkeywords}

\section{Introduction}

\IEEEPARstart{A}{udio-based} disease prediction, focused on deducing pathological symptoms through human acoustic bio-signals such as cough, breathing, and speech, has become a trending research area \cite{cai2023discovering,xia2022exploring}. Leveraging deep learning algorithms, audio-based prediction systems have proven effective in a diverse array of disease diagnosis scenarios, such as respiratory ailments (both acute and chronic), mental health disorders, and developmental abnormalities \cite{schuller2021covid,shi2018theory,rohanian2020multi}. Benefiting from its non-invasive, cost-effective, and accessible nature, audio-based disease prediction could serve as a promising complement to traditional medical diagnostic tools. 



Due to the high-dimensional and noise-sensitive nature of raw audio clips, most audio-based disease prediction systems tend to extract and utilize features from various domains and sub-domains, such as time, frequency, and cepstral domains, rather than inputting raw data directly \cite{moran2022review,azadi2021robust}. These heterogeneous acoustic features are mappings of a specific bio-acoustic modality within different dimensional spaces, revealing various aspects of its characteristics for disease diagnosis.  Moreover, even identical feature types from various bio-acoustic modalities, such as coughs and breath sounds, can offer valuable insights into unique facets of disease symptoms.

\textcolor{black}{Drawing parallels from these insights, multimodal fusion methods that involve merging acoustic features from different domains within a single bio-acoustic modality \cite{fusefeature1,fusefeature3,chen2024c3,campana2023transfer,ulukaya2023msccov19net,de2023covid}, i.e., intra-modal fusion, or combining acoustic features across multiple modalities \cite{dang2022exploring,liu2021covid,mallol2022multi,krstev2022multimodal,celik2023covidcoughnet,harvill2022detection,pentakota2023screening,effati2023performance,alvarado2023dyspnea,skaramagkas2023using}, i.e., inter-modal fusion,} have been developed to improve disease prediction outcomes compared to unimodal methods. Despite recent promising results, several key challenges should be surmounted to fully harness the potential of multimodal audio-based disease prediction, as outlined below. 

\textcolor{black}{\textit{Unilateral Fusion Strategies.}} Most existing studies exclusively adopt either intra-modal \cite{fusefeature1,fusefeature3,chen2024c3,campana2023transfer,ulukaya2023msccov19net,de2023covid} or inter-modal \cite{dang2022exploring,liu2021covid,mallol2022multi,krstev2022multimodal,celik2023covidcoughnet,harvill2022detection,effati2023performance,pentakota2023screening,alvarado2023dyspnea,skaramagkas2023using} fusion, rarely exploring their simultaneous application. \textcolor{black}{While intra-modal fusion methods can capture a broad range of characteristics within a specific bio-acoustic modality by fusing features extracted from different domains, they often miss the synergistic benefits achievable through integrating multiple modalities. On the other hand, while inter-modal methods can provide such benefits, they may overlook the deep, nuanced interconnections across diverse feature domains within each modality, since they often utilize features from a single domain for each modality \cite{dang2022exploring,mallol2022multi,effati2023performance} or simply concatenate \cite{krstev2022multimodal,pentakota2023screening} or average \cite{celik2023covidcoughnet} several features of one modality. In summary, the prevalent unidirectional fusion pattern may limit the model to fully exploit the complementary information derived from various fusion stages.} To address this deficiency, it is imperative to explore a comprehensive fusion strategy that effectively combines the fusion processes within and across bio-acoustic modalities.

\textcolor{black}{\textit{Inadequate Latent Dependencies Exploration.}} Bio-acoustic features and modalities are inherently heterogeneous yet latently complementary, each offering unique insights into diverse patterns crucial for disease diagnosis. \textcolor{black}{To learn efficient unimodal and fused representations that leverage such incongruity across different feature domains and bio-acoustic modalities during intra- and inter-modal fusion, it is essential to explore the latent dependencies in modality-specific and modality-shared spaces. While such explorations have been well-studied in fields like computer vision and natural language processing \cite{xu2023multimodal}, the context of audio-based disease prediction remains unexplored. Most current works in this field rely on simple alignments and concatenations to fuse features either within \cite{fusefeature1} or across modalities \cite{liu2021covid,dang2022exploring,de2023covid}, or they process each feature or modality individually through score-level or decision level fusion \cite{chen2024c3,fusefeature3,campana2023transfer,effati2023performance,alvarado2023dyspnea}. While recent studies \cite{mallol2022multi, truong2024fused} have employed certain attention mechanisms or correlation analysis methods to learn shared weights or representations, they often fail to capture intra- and inter-modal dependencies simultaneously and comprehensively.}

\textit{Limited Applicability in General Scenarios.} Different bio-acoustic features and modalities demonstrate varying degrees of sensitivity and effectiveness in relation to different diseases and task scenarios. Given this variability, most existing studies \cite{dang2022exploring,liu2021covid,mallol2022multi} employ meticulous feature selection processes, with the goal of customizing their models to achieve high performance in specific task settings. However, this approach necessitates extensive prior knowledge and cross-validation to be effective, which inherently limits the applicability across a wider range of diseases and scenarios. Therefore, existing models, often designed and validated for a specific disease or a certain combination of features, may not serve as robust backbone networks for general audio-based disease prediction.

\begin{figure*}[t]
    \centering
    \includegraphics[width=\linewidth]{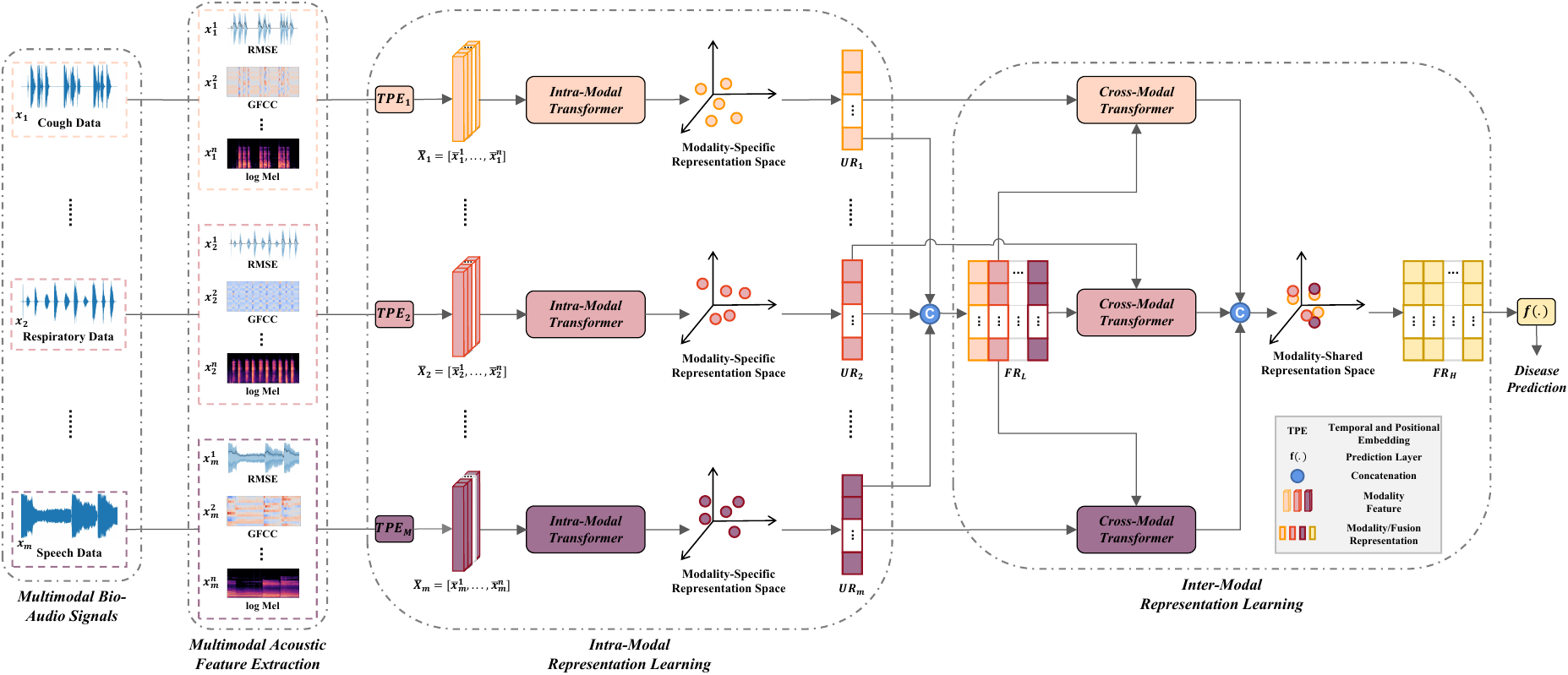}
    \vspace{-15pt}
    \caption{\textcolor{black}{Illustration of the proposed \textit{AuD-Former} framework.} This illustration showcases the framework using cough, respiration, and speech modalities as example inputs; however, the framework is versatile and can accommodate a variety of bio-audio modalities. Initially, multimodal low-level acoustic features extracted from multiple bio-audio sources undergo temporal and positional embedding processes, resulting in sequences of temporal unimodal features denoted as $\overline{X}_{{1, \cdots, m}}$ (see Section \ref{TE}). These sequences are input into an intra-modal representation learning module composed of multiple intra-modal transformer networks. This module produces unimodal representations $\textit{UR}_{{1, \cdots, m}}$, which effectively capture intra-modal dependencies within each modality-specific context (see Section \ref{Intra}). Subsequently, these unimodal representations are concatenated and, along with a low-level fusion representation $\textit{FR}_L$, fed into an inter-modal representation learning module. This module constructs a high-level fusion representation $\textit{FR}_H$ that encodes latent cross-modal complementarities within a shared modality space (see Section \ref{Inter}). Finally, the high-level fusion representation $\textit{FR}_H$ passes through a prediction layer, consisting of a multi-head attention sub-layer followed by two linear sub-layers, to produce the disease prediction.}
    \vspace{-10pt}
    \label{fig:framework}
\end{figure*}

In response to these challenges, \textcolor{black}{we introduce a transformer-based hierarchical fusion network, named \textit{AuD-Former}, for general multimodal audio-based disease prediction} as illustrated in Fig. \ref{fig:framework}. The primary contributions of this work can be summarized as:

\begin{itemize}[leftmargin=*]
    \item \textcolor{black}{We propose a hierarchical fusion strategy to emphasize both intra-modal and inter-modal fusion for multimodal audio-based disease prediction tasks, effectively exploiting the complementary nature of different feature domains within and across bio-acoustic modalities.}
    \item  To adequately capture dependencies within both modality-specific and modality-shared spaces, we introduce intra-modal and inter-modal representation learning modules. \textcolor{black}{This approach allows the hierarchical fusion to query an informative multimodal representation using unimodal features,} thus eliminating the need for the meticulous feature selections common in previous works and enhancing the overall generalizability of our model as a robust backbone network.
    \item \textcolor{black}{Our extensive evaluations, conducted on five datasets across three distinct diseases: COVID-19, Parkinson's disease, and pathological dysarthria, demonstrate that our model surpasses existing state-of-the-art multimodal fusion methods in the audio-based disease prediction}. Additionally, ablation studies and qualities analysis further investigate the contributions of the main components within the \textit{AuD-Former} framework, showing their individual and combined impacts.
\end{itemize}

\section{Background and Related Works}

\textcolor{black}{In this work, we define a \textit{modality} as a distinct type of vocal behavior or bio-acoustic signal (e.g., cough, breathing, speech) generated by activation of different body parts, including larynx, vocal folds, tongue, lips, and palate, each offering unique insights into a patient’s health status. We use the term \textit{multimodal fusion} to describe the process of integrating these various audio modalities or their different feature domains to form a comprehensive representation \cite{bleiholder2009data}.} In the context of audio-based disease prediction, a common practice is to extract features from various domains like time, frequency, and cepstral, from raw audio clips \cite{moran2022review}. This characteristic introduces two types of multimodal fusion in literature: the fusion of different bio-acoustic modalities, such as cough, breathing, and speech (known as inter-modal fusion), and the fusion of different feature domains within a single modality (referred to as intra-modal fusion).

\textcolor{black}{A substantial body of research emphasizes inter-modal fusion, involving the integration of multimodal representations across various modalities \cite{dang2022exploring,liu2021covid,mallol2022multi,krstev2022multimodal,celik2023covidcoughnet,harvill2022detection,pentakota2023screening,skaramagkas2023using, truong2024fused} and the combination of insights from models trained on individual modalities \cite{9746205,effati2023performance,alvarado2023dyspnea,liu2022automatic,liu2019acoustical}. However, these approaches often neglect the rich intra-modal correlations as they generally utilize a single pre-trained model or method for feature extraction within each modality, typically focusing on a limited set of feature domains. For instance, Dang \textit{et al.} \cite{dang2022exploring} employed pre-trained VGGish \cite{hershey2017cnn} models to independently extract unimodal representations for cough, breathing, and voice sounds, which were then concatenated and input into a GRU \cite{chung2014empirical} network for COVID-19 prediction. This method potentially overlooks valuable insights from other feature domains within each modality. Furthermore, latent inter-modal dependencies may not be fully captured due to limited consideration of the complex interactions between different bio-acoustic signals. For example, Effati \textit{et al.} \cite{9746205} implemented shared weight strategies to synchronize knowledge across modalities by averaging weights among three BiLSTM models, each trained on specific data types. While this strategy aims to foster cross-modal integration, it may fall short in addressing the intricate relationships and dependencies due to its simplistic weight averaging mechanism.}

\textcolor{black}{On the other hand, several studies \cite{fusefeature1,fusefeature3,chen2024c3,campana2023transfer,ulukaya2023msccov19net,de2023covid} have concentrated on intra-modal fusion. However, these approaches often confine their methods within a single modality without integrating inter-modal fusion. Moreover, the exploration of intra-modal dependencies typically lacks depth: many opt for early concatenation that depends on aligning multiple features \cite{fusefeature1,de2023covid} or late score/decision-level fusion that processes each feature domain separately \cite{chen2024c3,fusefeature3,campana2023transfer,liu2019acoustical}. For instance, Bhosale \textit{et al.} \cite{fusefeature1} utilized the concatenation of multiple temporal, spectral, and tempo-spectral features as input to an early fusion model for COVID-19 detection. Additionally, Liu \textit{et al.} \cite{liu2019acoustical} developed two MLP classifiers, each tailored to specific feature sets, with their classification scores fused for the final prediction of voice disorders. These methods may fail to effectively facilitate communication between different feature domains due to the inadequate consideration of the intra-modal correlations.} 

\textcolor{black}{Additionally, to the best of our knowledge, no existing works have been developed for general audio-based disease prediction, proven to be effective across multiple diseases. Existing works  often design their models to specialize in specific combinations of features or modalities tailored for one particular disease, which limits their broader applicability.}

\section{methodology}
In this section, we present our proposed hierarchical transformer network for multimodal audio-based disease prediction.

\subsection{Problem Formulation and Framework Overview}
Consider multimodal audio signals composed of $m$ modalities. For each modality, the unimodal features extracted across $n$ different domains or subdomains can be represented as a low-level unimodal feature sequence $X_{(.)}=[x^1_{(.)}, x^2_{(.)}, \cdots, x^n_{(.)}] \in \mathbb{R}^{l_{(.)} \times d_{(.)}}$. In this paper, $l_{(.)}$ and $d_{(.)}$ denote the feature length and dimension of one modality, respectively. The classification task is to generate discrete labels for disease prediction based on these constituent multimodal audio features.


\textcolor{black}{Our scientific hypothesis is that a hierarchical two-step fusion strategy—first integrating features within individual modalities before combining across modalities—will more effectively capture the complementary relationships in audio-based disease indicators compared to conventional unilateral fusion approaches. We propose that this systematic progression from modality-specific to modality-shared representations, enhanced by appropriate attention mechanisms at each level, will enable more comprehensive feature integration for improved disease prediction performance.} To this end, we propose  \textit{AuD-Former}, a hierarchical transformer network designed to hierarchically capture sufficient intra-modality dependencies and inter-modality correlations, thereby providing an efficient acoustic fusion representation for downstream disease prediction tasks. 

As illustrated in Fig. \ref{fig:framework}, the \textit{AuD-Former} consists of two hierarchical core components: 1) \textit{Intra-modal representation learning}: \textcolor{black}{Utilizing intra-modal attention layers, we generate unimodal representations, denoted as $\textit{UR}$.} \textcolor{black}{These representations capture latent intra-modal correlations between various low-level features within a single modality, effectively mapping information across multiple domains} (Section \ref{Intra}); and 2) \textit{Inter-modal representation learning}: \textcolor{black}{Through inter-modal attention layers, we merge these heterogeneous unimodal representations into a unified fusion representation, denoted as $\textit{FR}$.} \textcolor{black}{This fusion effectively encodes cross-modal dependencies, allowing each target unimodal representation to continuously integrate complementary information from other modalities to enhance its own feature set} (Section \ref{Inter}). \textcolor{black}{These hierarchical modules are specifically designed to leverage the heterogeneity and latent complementary attributes within and across unimodal features of different modalities, overcoming the limitations of unilateral fusion strategies and inadequate dependency exploration in existing models.}

\subsection{Temporal and Positional Embedding}
\label{TE}
The low-level feature sequences of each modality, $X_{(.)}=[x^1_{(.)}, x^2_{(.)}, \cdots, x^n_{(.)}] \in \mathbb{R}^{l_{(.)} \times d_{(.)}}$, are first embedded by multiple 1-D temporal convolution (TC) layers to obtain convoluted unimodal feature sequences with the same dimension, $\hat{X}_{(.)} = [\hat{x}^1_{(.)}, \hat{x}^2_{(.)}, \cdots, \hat{x}^n_{(.)}] \in \mathbb{R}^{l_{(.)} \times d_{tc}}$ as:
\begin{equation}
\hat{X}_{\{1, \cdots, m\}} = \operatorname{TC}\left(X_{\{1, \cdots, m\}}, \Theta_{\{1, \cdots, m\}}\right)
\label{eqconv}
\end{equation}
\noindent where $\Theta_{\{1, \cdots, m\}}$ \textcolor{black}{represents the kernels of the temporal convolution layers, which have different sizes for various modalities.} These temporal convolution layers are designed to map heterogeneous unimodal features into a \textcolor{black}{$d_{tc}$}-dimensional homogeneous subspace. This process introduces time-related features and, more importantly, enables the dot-product operations in the following intramodal and intermodal representation learning modules to be mathematically feasible.

Furthermore, to account for the positional information of the unimodal sequence, we conduct triangle positional embeddings (PE) \cite{vaswani2017attention} to convoluted unimodal feature sequence to obtain temporal unimodal feature sequences $\overline{X}_{\{1, \cdots, m\}}\in \mathbb{R}^{l_{\{1, \cdots, m\}}, d_{tc}}$ as:
\begin{equation}
\overline{X}_{\{1, \cdots, m\}} = \operatorname{PE}\left(\hat{X}_{\{1, \cdots, m\}}\right)
\label{eqpe}
\end{equation}
\subsection{Intra-modal Representation Learning}
\label{Intra}

The unimodal features within a single modality originate from different domains or sub-domains, providing a unique perspective and emphasizing distinct characteristics of the modality. To capitalize on this heterogeneity and learn a comprehensive unimodal representation, we feed temporal unimodal feature sequence $\overline{X}_{\{1, \cdots, m\}}$ of each modality into the intra-modal transformer to generate unimodal representations with latent intra-modal correlations mined efficiently. 

The core of the intra-modal transformers is the multi-head self-attention mechanism \cite{vaswani2017attention}. Specifically, the self-attention process assesses pairwise relationships of each element in the unimodal feature sequence, i.e., the convoluted unimodal features obtained through temporal embedding, to integrate the contextual information from the entire feature sequence. Formally, we define queries $Q_{\{1, \cdots, m\}}$, keys $K_{\{1, \cdots, m\}}$ and values $V_{\{1, \cdots, m\}}$ for unimodal feature sequences as:
\begin{equation}
\label{self_qkv}
\begin{aligned}
Q_{\{1, \cdots, m\}} &= \overline{X}_{\{1, \cdots, m\}} \cdot W_{q_{\{1, \cdots, m\}}}\\
K_{\{1, \cdots, m\}} &= \overline{X}_{\{1, \cdots, m\}} \cdot W_{k_{\{1, \cdots, m\}}}\\
V_{\{1, \cdots, m\}} &= \overline{X}_{\{1, \cdots, m\}} \cdot W_{v_{\{1, \cdots, m\}}}
\end{aligned}
\end{equation}
\noindent where $W_{q_{\{1, \cdots, m\}}} \in \mathbb{R}^{d_{tc}, d_{sq}}$, $W_{k_{\{1, \cdots, m\}}} \in \mathbb{R}^{d_{tc}, d_{sk}}$, and $W_{v_{\{1, \cdots, m\}}} \in \mathbb{R}^{d_{tc}, d_{sv}}$ are three weight groups to be trained respectively \textcolor{black}{and $d_{sq}$ = $d_{sk}$}. 

Then the self-attention (SA) process is formulated as:
\begin{equation}
\begin{aligned}
\label{eq_sa}
\widehat{\textcolor{black}{\textit{UR}}}_{\{1, \cdots, m\}} &= {\operatorname{SA} (Q_{\{1, \cdots, m\}}, K_{\{1, \cdots, m\}}, V_{\{1, \cdots, m\}})}\\ &= \operatorname{softmax}\left(\frac{{Q_{\{1, \cdots, m\}}} \cdot {K_{\{1, \cdots, m\}}}^\top}{\sqrt{d_{sk}}}\right)\cdot{V_{\{1, \cdots, m\}}}
\end{aligned}
\end{equation}
\noindent where $\widehat{\textcolor{black}{\textit{UR}}}_{\{1, \cdots, m\}}  \in \mathbb{R}^{l_{\{1, \cdots, m\}}, d_{sv}}$ represent the unimodal representations resulting from single-head SA operation.
\begin{figure}[ht]
\centering
\includegraphics[width=0.63\columnwidth]{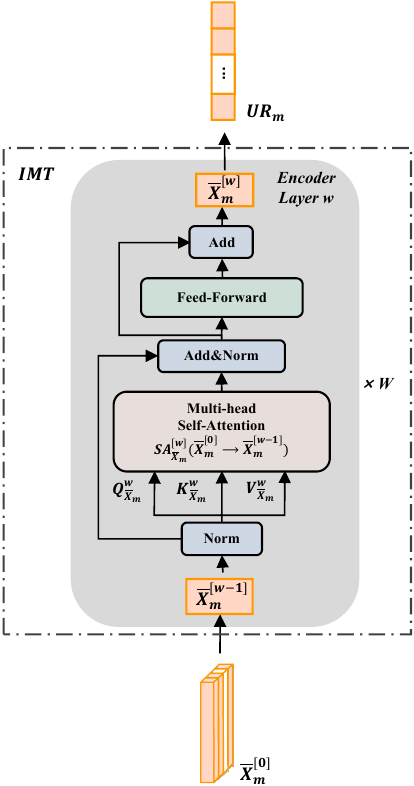}
\caption{Illustration of the intra-modal transformer network \textcolor{black}{for modality $m$.}}\label{fig:trans}
\vspace{-10pt}
\end{figure}

The above process can be conducted in parallel multiple times as multi-head self-attention. Ultimately, we derive the final unimodal representation $\textcolor{black}{{\textit{UR}}_{\{1, \cdots, m\}} \in \mathbb{R}^{l_{\{1, \cdots, m\}}, d_{sv}}}$ from $\widehat{\textcolor{black}{\textit{UR}}}_{\{1, \cdots, m\}}$ through multiple layer normalization and feed-forward operations within the intramodal transformer, as illustrated in Fig. \ref{fig:trans}. By computing different attention scores to unimodal features within a single bio-acoustic modality, the self-attention process adaptively accounts for interactions among various unimodal features in modality-specific spaces, effectively encoding the complementary information they provide into the unimodal representation.

\subsection{Inter-modal Representation Learning}
\label{Inter}
In practical situations, medical professionals must thoroughly examine and combine clinical data from various sources to make well-founded diagnostic decisions. Likewise, a dependable multimodal diagnostic system needs to be proficient at leveraging the commonalities and complementarities across different bio-acoustic modalities. Typically, commonalities of multiple modalities are thought to reflect consistent information about the disease, whereas complementarities convey supplementary information. To this end, we propose the inter-modal representation learning module to effectively mine adequate complementary dependencies and adaptations across different modalities. 

\begin{figure}[t]
\centering
\subfloat[Procedures of the cross-modal attention mechanism.\label{fig:cm}]{\includegraphics[width=0.8\columnwidth]{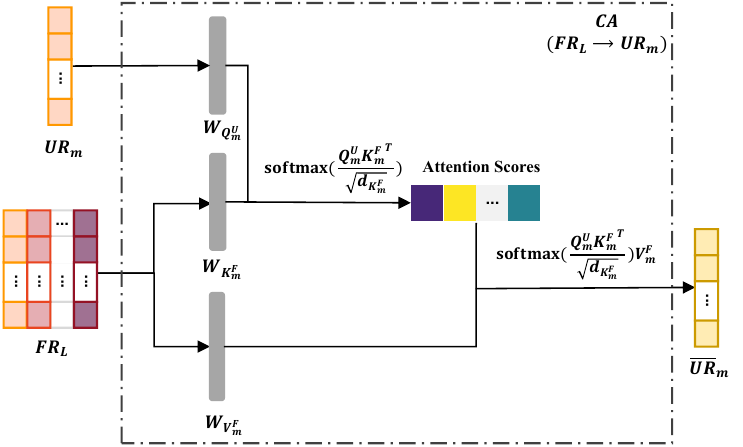}}

\subfloat[Architecture of the cross-modal transformer network.\label{fig:cmt}
]{\includegraphics[width=0.7\columnwidth]{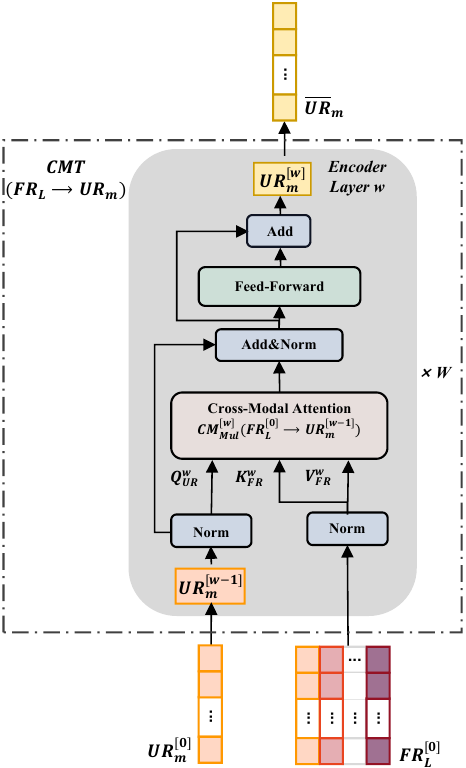}}
\caption{Illustration of the cross-modal attention (CA) mechanism and cross-modal transformer network.}
\vspace{-15pt}
\label{fig:c}
\end{figure}

The unimodal representations of all modalities ${\textcolor{black}{\textit{UR}}}_{\{1, \cdots, m\}}$ first produce the low-level fusion representation ${\textcolor{black}{\textit{FR}}}_L  \in \mathbb{R}^{l_f, d_{sv}}$ with the concatenation operation. This representation is then fed, along with each unimodal representation respectively, into multiple cross-modal transformers. Each cross-modal transformer aims to progressively enhance the target unimodal representation ${\textcolor{black}{\textit{UR}}}_m$ with other modalities encoded in the fusion representation ${\textcolor{black}{\textit{FR}}}_L$ by computing the cross-modal attention illustrated in Fig. \ref{fig:cm}. We formally define queries \textcolor{black}{$Q^{U}_{\{1, \cdots, m\}}$} derived from the target unimodal representations, and keys $K^{F}{{1, \cdots, m}}$ and values $V^{F}_{{1, \cdots, m}}$ derived from the source fusion representations as:
\begin{equation}
\label{cross_qkv}
\begin{aligned}
Q^{U}_{\{1, \cdots, m\}} &= \textit{{UR}}_{\{1, \cdots, m\}} \cdot W_{q^{U}_{\{1, \cdots, m\}}}\\
K^{F}_{\{1, \cdots, m\}} &= {\textcolor{black}{\textit{FR}}}_L \cdot W_{k^{F}_{\{1, \cdots, m\}}}\\
V^{F}_{\{1, \cdots, m\}} &= {\textcolor{black}{\textit{FR}}}_L \cdot W_{v^{F}_{\{1, \cdots, m\}}}
\end{aligned}
\end{equation}
where $W_{q^{U}_{\{1, \cdots, m\}}} \in \mathbb{R}^{d_{sv}, d_{cq}}$, $W_{k^{F}_{\{1, \cdots, m\}}} \in \mathbb{R}^{d_{sv}, d_{ck}}$, and $W_{v^{F}_{\{1, \cdots, m\}}} \in \mathbb{R}^{d_{sv}, d_{cv}}$ denote three trainable weights respectively \textcolor{black}{and $d_{cq}$ = $d_{ck}$}. \textcolor{black}{These matrices enable the model to adapt and transform features for effective cross-modal information exchange.}

Correspondingly, the cross-modal attention (CA) process is denoted as:
\begin{equation}
\begin{aligned}
\label{eq_ca}
\dot{\textcolor{black}{\textit{UR}}}_{\{1, \cdots, m\}} &= {\operatorname{CA} (Q^{U}_{\{1, \cdots, m\}}, K^{F}_{\{1, \cdots, m\}}, V^{F}_{\{1, \cdots, m\}})}\\ &= \operatorname{softmax}\left(\frac{Q^{U}_{\{1, \cdots, m\}} \cdot {K^{F}_{\{1, \cdots, m\}}}^\top}{\sqrt{d_{ck}}}\right)\cdot{V^{F}_{\{1, \cdots, m\}}}
\end{aligned}
\end{equation}
\noindent where $\dot{\textcolor{black}{\textit{UR}}}_{\{1, \cdots, m\}}  \in \mathbb{R}^{l_{\{1, \cdots, m\}}, d_{cv}}$ represent the outputs of the single-head cross-attention operation.

\textcolor{black}{This process encourages each unimodal representation ${\textcolor{black}{\textit{UR}}}_m$ to attend to other unimodal representations within ${\textcolor{black}{\textit{FR}}}_L$, learning significant complementarities and commonalities to reinforce itself. The external complementary information from other modalities is encoded into multiple fusion keys \textcolor{black}{$K^{F}_{\{1, \cdots, m\}}$} and values \textcolor{black}{$V^{F}_{\{1, \cdots, m\}}$}, guiding adaptations to the target modality through inter-modal attention.} This procedure is executed concurrently several times as multi-head cross-modal attention.

Subsequently, we obtain enhanced unimodal representations \textcolor{black}{$\overline{\textit{UR}}_{\{1, \cdots, m\}}$} from \textcolor{black}{$\dot{\textit{UR}}_{\{1, \cdots, m\}}$} via multiple layer normalization and feed-forward operations, as depicted in Fig. \ref{fig:cmt}. Finally, all reinforced unimodal representations are combined to derive the high-level fusion representation $\textcolor{black}{\textit{FR}_H} \in \mathbb{R}^{l_f, d_{cv}}$ in the modality-shared space for downstream disease prediction.

\subsection{Prediction Layer and Model Optimization}
\label{Model}
To further distill essential contextual information for disease diagnosis, the representation $\textcolor{black}{\textit{FR}}_H$ is additionally processed through a layer featuring multi-head self-attention, as depicted in Eq. \ref{eq_sa}. The output, denoted as $\overline{\textcolor{black}{\textit{FR}}}_H \in \mathbb{R}^{l_f, d_{cv}}$, is then put into subsequent linear layers accompanied by residual operations and featuring softmax activation functions to generate disease predictions, formally defined as:
\begin{equation}
\begin{split}
\hat{\textcolor{black}{\textit{FR}}}_H &= \overline{\textcolor{black}{\textit{FR}}}_H + \varrho_\nu(\overline{\textcolor{black}{\textit{FR}}}_H)\\  
\mathcal{P} &= \operatorname{softmax}(\varrho_\tau(\hat{\textcolor{black}{\textit{FR}}}_H)\\
\hat{y} &= \operatorname{argmax}(\mathcal{P}^j)
\label{p}
\end{split}    
\end{equation}
\noindent where $\hat{y}\in \mathbb{R}^1$, \textcolor{black}{with $\hat{y}_i \in \{0,1\}$ }and $\mathcal{P}_j \in \mathbb{R}^2$ represent the predicted labels and probabilities for the $j_{th}$ class (two classes in our setting: Positive or Negative) in the disease prediction task, and $\varrho_\nu$ and $\varrho_\tau$, denote two fully-connected layers with parameter sets $\nu$ and $\tau$, respectively.

For model optimization, we choose binary cross-entropy loss defined as:
\begin{equation}
\mathcal{L}(y, \hat{y})=-\frac{1}{N} \sum_{i=1}^N\left[y_i \log \left(\hat{y}_i\right)+\left(1-y_i\right) \log \left(1-\hat{y}_i\right)\right]\\
\end{equation}
\noindent where ${y_i}$ and ${\hat{y}_i}$ are the ground-truth and predicted labels for the $i^{th}$ instance, respectively.

\section{Experimental Setting}
\textcolor{black}{In this section, we detail the experimental setup used to evaluate our proposed \textit{AuD-Former}. The core objective is to address the research question: \textit{Can the hierarchical integration of intra-modal and inter-modal fusion processes enable the efficient querying of multimodal representations using unimodal feature sets, thereby enhancing the performance of general audio-based prediction tasks?}}

\textcolor{black}{To this end, we compared the \textit{AuD-Former} to other state-of-the-art audio-based prediction baselines that utilize intra-modal or/and inter-modal fusion across various diseases, such as respiratory diseases, neurological disorders, and speech disorders. We also extensively implemented extra baselines and ablation models to investigate the contributions of the main modules within the \textit{AuD-Former} framework, such as intra-modal and inter-modal representation learning, and the hierarchical fusion strategy in addition to the network itself.}

\subsection{Dataset Description}
\textcolor{black}{We evaluated the \textit{AuD-Former} using five datasets, focusing on prediction of three distinct diseases: COVID-19, Parkinson's disease (PD), and pathological dysarthria.} 

For COVID-19 classification, we selected two datasets. (1) Coswara Dataset \cite{sharma2020coswara}: a large-scale benchmark dataset featuring audio recordings collected during the COVID-19 pandemic. It encompasses bio-acoustic data from four primary modalities: breathing (deep and shallow), coughing (heavy and shallow), counting (fast and slow), and vowel speech modalities (including /\textscripta/, /i/, /u/). The dataset includes recordings from 2,635 subjects categorized according to their self-reported health conditions: 674 individuals tested positive for COVID-19 (asymptomatic, mild, and moderate symptoms), 1,819 individuals reported as healthy or suffering from another respiratory illness, and 142 individuals who have fully recovered from COVID-19. Following established protocols \cite{truong2024fused,9746205,muguli2021dicova}, we classified instances from subjects with mild and moderate symptoms as COVID-19 Positive and those from the healthy category as Healthy. \textcolor{black}{(2) Sound-Dr dataset \cite{nguyen2023sound}: a high-quality human sound dataset aimed at respiratory disease detection. It includes recordings from three modalities: mouth breathing, nose breathing, and coughing. Following the approach outlined in \cite{nguyen2023sound}, we classified patients who tested positive for COVID-19 within the last 14 days as COVID-19-positive and all others as Healthy.}

For Parkinson's disease classification, we utilized two popular datasets. (1) IPVS dataset \cite{italian_dataset}. It consists of three pronunciation recording modalities including phonetically balanced text reading, phonetically balanced phrases reciting, and syllables /p\textscripta/ and /t\textscripta/ pronunciation.\textcolor{black}{(2) PC-GITA dataset \cite{orozco2014new}, a Spanish speech corpus designed for PD classification. For our experiments, we selected six phrase recordings (apto, drama, gato, grito, ñame, reina) as the phrase reading modality, /p\textscripta k\textscripta t\textscripta/ and /p\textscripta t\textscripta k\textscripta/ recordings as the diadochokinetic (DDK) modality, three sentence recordings (begin with Viste, Luisa, Rosita) as the sentence reading modality, and the vowel /\textscripta/ recording as the vowel modality \cite{vasquez2018towards,orozco2015spectral}. Instances from these datasets were classified into PD-Positive and Healthy classes separately. }

\textcolor{black}{We further evaluate our model on the Saarbruecken Voice Database (SVD) \cite{woldert2007saarbruecken}, which is a benchmark dataset for pathological dysarthria. It contains modalities of phrases and vowels (/\textscripta/, /i/, /u/) of high, neutral, and low pitch, from Pathological Dysarthria patients and Healthy controls. } 

An overview of the data distribution for these datasets in terms of male, female, positive, and negative cases is provided in Table~\ref{table:datasets}.

\begin{table}[t]
\centering
\caption{\textcolor{black}{Distribution of the utilized data on all datasets.} }
\label{table:datasets}

\begin{tabular}{cccccc}
\hline
\multirow{2}{*}{Dataset} & \multicolumn{2}{c}{Positive}      & \multicolumn{2}{c}{Negative}       \\ \cline{2-5} 
                         & \multicolumn{1}{c}{Male} & Female & \multicolumn{1}{c}{Male} & Female \\ \hline\hline
                    
Coswara \cite{sharma2020coswara}                 & \multicolumn{1}{c}{212}  & 138    & \multicolumn{1}{c}{1086} & 373    \\ 
Sound-Dr \cite{nguyen2023sound}                 & \multicolumn{1}{c}{143}  & 112    & \multicolumn{1}{c}{491} & 285    \\
IPVS \cite{italian_dataset}                     & \multicolumn{1}{c}{19}   & 9      & \multicolumn{1}{c}{10}   & 12     \\ 
PC-GITA \cite{orozco2014new}                 & \multicolumn{1}{c}{25}  & 25    & \multicolumn{1}{c}{25} & 25    \\
SVD \cite{woldert2007saarbruecken}                 & \multicolumn{1}{c}{452}  & 559    & \multicolumn{1}{c}{252} & 377    \\ \hline

\end{tabular}
\vspace{-15pt}
\end{table}

\subsection{Data Pre-processing and Feature Extraction}

We first preprocess our dataset by concatenating multiple recordings under the same modality for each subject. For instance, deep and shallow breathing clips are combined to form a single breathing modality data set for the Coswara dataset, and high, neutral, and low pitch vowel sounds are merged to form the vowel modality for the SVD dataset. Following previous work \cite{9746205}, we excised silent portions, resulting in more concise and relevant audio segments. We then standardized the audio clips to a uniform length across all subjects for each modality, accommodating the varying durations of the original recordings. After preprocessing, we further standardized audio clips to specific lengths based on their respective modalities. \textcolor{black}{For Coswara, we standardized cough clips at 8 seconds, breathing clips at 19 seconds,
counting clips at 18 seconds, and vowel clips at 20 seconds; For Sound-Dr, we standardized audio clips of all modalities at 15 seconds; For IPVS, clips for all modalities are standardized at 5 seconds; For PC-GITA, we standardized phrase reading clips at 3 seconds, sentence reading clips at 15 seconds,
DDK and vowel clips both at 6 seconds; For SVD, clips of all three vowels are standardized at 6 seconds, while phrase modality clips are standardized at 3 seconds.}

We extracted features from seven commonly used domains for audio classification tasks, as outlined in \cite{sharma2020trends}: Zero Crossing Rate (ZCR), Short-Time Energy (STE), Spectral Centroid (SC), Log-Mel Spectrogram, Mel Frequency Cepstral Coefficients (MFCC), GammaTone Frequency Cepstral Coefficients (GFCC), and Constant Q Cepstral Coefficients (CQCC). GFCC features were extracted using the \textit{Spafe} library, while the remaining features were obtained with the \textit{librosa} library, both at a standard sampling rate of $44.1$ kHz. Detailed descriptions of the feature extraction processes are available at our website: \url{https://sites.google.com/view/audformer}. After extracting these features, each input instance comprises a set of $7*n$ features from $n$ modality groups. Additionally, following previous works \cite{pahar2021covid,baseline1}, we applied the Synthetic Minority Over-sampling Technique (SMOTE) \cite{chawla2002smote} to datasets exhibiting extreme imbalance, specifically the Coswara and Sound-Dr.

\subsection{Baselines and Ablation Models}
We selected and presented the results of several state-of-the-art baselines that utilize either intra-modal or inter-modal fusion (or both) for audio-based disease prediction, serving as comparisons to our proposed \textit{AuD-Former}:

\begin{itemize}[leftmargin=*]
    \item \textcolor{black}{\textit{AE+RF} \cite{tena2022automated}: This model utilizes the Random Forest (RF) method, which is trained on 15 features extracted from cough audio using an Autoencoder (AE). We compare our model's performance against its C vs. NCC setting, which matches our dataset configuration.} It serves as a baseline for intra-modal fusion on the Coswara dataset.
    \textcolor{black}{\item \textit{FRILL+SVM} \cite{nguyen2023sound}: This model employs the pre-trained FRILL model, which is based on the MobileNet architecture, for audio feature extraction. The features extracted are then classified using SVMs with linear kernels. This setup serves as the benchmark performance of the Sound-Dr.}
    \item \textit{CNN-EMD} \cite{baseline4}: This model utilizes Empirical Mode Decomposition (EMD) to extract features from vowel sounds, processed through multiple 1D-CNN layers. Features are concatenated and used for PD prediction, serving as an intra-modal fusion baseline on the IPVS.
    \item \textit{U-Lossian} \cite{maskeliunas2022hybrid}: This model features a hybrid Mask U-Net architecture with adaptive custom loss functions, extracting local and global features of the speech modality integrated via skip connections for PD prediction. It denotes an intra-modal fusion baseline on the IPVS.
    \textcolor{black}{\item \textit{NCA+SVM} \cite{reddy2021comparison}: This model utilizes a neighborhood component analysis (NCA) feature selection technique to extract combined MFCC features from source-based and vocal tract-based cepstral characteristics of vowel /\textscripta/ sounds. Following feature selection, it employs an SVM with a radial basis function (RBF) kernel for classification. This method serves as an intra-modal fusion baseline on the PC-GITA.}
    \textcolor{black}{\item \textit{QCP Glottal flow} \cite{narendra2021detection}: This model integrates multiple layers of CNN and MLP to analyze the glottal flow wave, which is extracted using quasi-closed phase (QCP) glottal inverse filtering techniques. It processes various continuous speech modalities, including DDK exercises, reading phrases and sentences, and delivering monologues. Long speech clips are divided into uniform-length segments, and scores from each segment are averaged to perform final binary classification. This presents an inter-modal fusion baseline on the PC-GITA.}
    \textcolor{black}{\item \textit{DW+CLL+CNL} \cite{zhang2023robust}: This model designed a framework for feature embedding extraction for dysphonic voice detection. It employs data-warping (DW) techniques to augment the original data, which is then processed by an encoder for contrastive loss (CNL) and an MLP classifier for classification loss (CLL). CNL and CLL are combined to jointly train the model, enhancing its learning efficiency. The features are specifically extracted from vowel /\textscripta/ sounds. This model serves as an intra-modal fusion baseline on the SVD dataset.}
    \textcolor{black}{\item \textit{Resnet18+SVM} \cite{lee2024quantitative}: This model combines features extracted from spectrograms using Resnet18 with handcrafted audio features generated by the OpenSmile toolkit from the phrase reading modality. The concatenated features are then classified using an SVM with RBF kernels. This approach serves as an intra-modal fusion baseline on the SVD dataset.}
    
\end{itemize}
\textcolor{black}{We also re-implemented two state-of-the-art baselines originally on Coswara for all datasets:}
\begin{itemize}[leftmargin=*]
    \item \textit{MM-Score} \cite{baseline3}: Multi-Modal Score-level Fusion model \textcolor{black}{combines various modalities by processing feature sets extracted from each modality through individual LSTM layers to generate prediction scores. A score-averaging scheme is then applied to produce the final predictions. This model serves as a baseline for inter-modal fusion.}
    \item \textit{FAIR} \cite{truong2024fused}: This model integrates spectral and waveform features from different audio modalities using \textit{DeiT-S/16} \cite{touvron2021training} and \textit{wav2vec} \cite{baevski2020wav2vec} encoders, which are fused via a multi-head self-attention layer for final prediction. This represents a baseline involving both intra- and inter-modal fusion.
\end{itemize}
Furthermore, to explore the benefits of the hierarchical fusion strategy beyond the hierarchical transformer network, we also implemented two extra baselines on all datasets:
\begin{itemize}[leftmargin=*]
    \item \textit{IntraFusion}: We implemented two advanced attention-based networks: Graph Attention Network (GAT) \cite{velivckovic2017graph} and Transformer Network \cite{vaswani2017attention}. Each was tested using the same feature inputs within each modality that we utilized in the \textit{AuD-Former}. Note that we employed a fully connected adjacency matrix for the GAT, assuming that each unimodal feature shares dependencies with one another. The best results achieved by these two networks on each modality are presented as the representative performance of the \textit{IntraFusion} model.
    \item \textcolor{black}{\textit{InterFusion}: This model utilizes a single feature domain within each bio-acoustic modality while maintaining the same hierarchical transformer network structure used in the \textit{AuD-Former}. The optimal results from various feature domains within each modality are reported to represent the performance of \textit{InterFusion}.}
\end{itemize}

Moreover, to investigate the benefits of the hierarchical transformer network brought to the hierarchical fusion strategy, we implemented two benchmark multimodal fusion models for time-series data \cite{lindemann2021survey,10413204}, serving as baselines utilizing the same multimodal inputs as the \textit{AuD-Former} on all datasets:

\begin{itemize}[leftmargin=*]
\item \textit{EF-LSTM}: \textcolor{black}{Long Short-Term Memory (LSTM) with early fusion. It involves concatenating the TC-processed multimodal features from different modalities, which are then input into an LSTM network. The final hidden state of the LSTM is used as the sequence encoding and passed through a classification layer to produce the final prediction.}
\item \textit{LF-LSTM}: LSTM with late fusion. \textcolor{black}{The TC-processed multimodal features of each modality are processed separately by individual LSTM networks. The final hidden states from these modality-specific LSTMs are concatenated and input into a final LSTM layer. The final hidden state of this last LSTM layer is used for generating predictions.}
\end{itemize}


To validate the benefit of each representation learning part inside the \textit{AuD-Former}, two ablation models are constructed: 
\begin{itemize}[leftmargin=*]
    \item \textit{IntraAtt}: \textcolor{black}{This model retains only the intra-modal representation learning module. After processing through intra-modal transformers, the resulting unimodal representations are directly concatenated and fed into the final prediction layer, bypassing inter-modal fusion}.
    \item \textit{InterAtt}: \textcolor{black}{This model removes the intra-modal representation learning module. The temporally encoded features from each modality are processed directly by cross-modal transformers, where individual modality features serve as queries while concatenated multimodal features serve as keys and values for attention computation.}
\end{itemize}

\subsection{Evaluation Scheme and Metrics}
We performed a random shuffle of all instances and conducted 10-fold cross-validation for each model on each dataset. Additionally, to evaluate the applicability of our model to new patients under realistic scenarios, we meticulously partitioned the instances from each individual into either the training set or the test set. In terms of evaluation metrics, we followed previous works \cite{baseline1,baseline3,truong2024fused} to report the average and standard deviation of Accuracy (ACC), F1 score, Area Under Curve (AUC), Sensitivity (SEN), and Specificity (SPE) of each model during experiments. 






\subsection{Implementation Details}

Experiments were conducted on an NVIDIA GeForce RTX 4090 GPU. In the \textit{EF-LSTM} model, we used temporal convolution layers, identical to the \textit{AuD-Former}, before the LSTM layer, enabling the mathematical feasibility of unimodal feature concatenation. To guarantee fair comparisons, hyper-parameters of ablation models remained consistent with those in the \textit{AuD-Former} during each run, which are available in Appendix. \ref{appendix:A}. The source code, along with detailed experimental details, can also be found on the project website.

\begin{table*}[t]
\centering
\caption{Summary of experimental results on the COVID-19 (Coswara, Sound-Dr) and Parkinson Disease (IPVS, PC-GITA) datasets in terms of average and standard deviations of Accuracy (ACC), F1 score, Area Under Curve (AUC), Sensitivity (SEN), and Specificity (SPE). \textit{EM}: Evaluation Method; $^h$: higher means better; $^*$: reported from literature;$\vartriangle$: re-implemented; -: not reported.  }
\label{table:baselines}
 \resizebox{\linewidth}{!}{
\begin{tabular}{c||cccccc||cccccc}
\hline
Dataset                       & \multicolumn{6}{c||}{Coswara}                                                                                                                                                                                        & \multicolumn{6}{c}{\textcolor{black}{Sound-Dr}}                                                                                                                                                                                           \\
Metric                        & \textit{EM} & \textit{ACC(\%$^h$)} & \textit{F1(\%$^h$)} & \textit{AUC(\%$^h$)} & \textit{SEN(\%$^h$)} & \textit{SPE(\%$^h$)} & \textit{EM}   & \textit{ACC(\%$^h$)} & \textit{F1(\%$^h$)} & \textit{AUC(\%$^h$)} & \textit{SEN(\%$^h$)} & \textit{SPE(\%$^h$)} \\ \hline
\textcolor{black}{\textit{AE+RF$^*$}}    & 10-fold           & \textcolor{black}{79.74}                                    & \textcolor{black}{79.52}                                   & \textcolor{black}{83.57}                                   & \textcolor{black}{79.70}                                   & \textcolor{black}{79.79}                                    & -                    & -                                    & -                                   & -                                    & -                                    & -                                    \\

\textit{FRILL+SVM$^*$}                            & -                 & -                                    & -                                   & -                                    & -                                    & -                                    & 5-fold               & 82.53                       & 70.48                      & 81.37$\pm$0.85                       & -                       & -                       \\\hline
\textcolor{black}{\textit{MM-Score$\vartriangle$}     }    & 10-fold            & \textcolor{black}{ 76.72$\pm$1.56                                    } & \textcolor{black}{ 75.82$\pm$1.85                               } & \textcolor{black}{ 76.30$\pm$1.55                                } & \textcolor{black}{ 75.82$\pm$1.13                                } & \textcolor{black}{ 76.78$\pm$2.85     } & \textcolor{black}{ 10-fold                                } & \textcolor{black}{ 83.65$\pm$3.76                    } & \textcolor{black}{ 78.64$\pm$2.81                                    } & \textcolor{black}{ 83.50$\pm$2.77                                   } & \textcolor{black}{ 79.59$\pm$3.29                                    } & \textcolor{black}{ 86.62$\pm$2.94  }                                                                     \\
\textcolor{black}{\textit{FAIR$\vartriangle$}  }       & 10-fold            & \textcolor{black}{ 78.74$\pm$5.46                                    } & \textcolor{black}{ 78.54$\pm$5.06                                   } & \textcolor{black}{ 78.98$\pm$1.15                       } & \textcolor{black}{ 78.57$\pm$5.54                       } & \textcolor{black}{ 79.58$\pm$6.78                       } & \textcolor{black}{ 10-fold                    } & \textcolor{black}{ 85.35$\pm$3.73                                    } & \textcolor{black}{ 84.24$\pm$3.41                                   } & \textcolor{black}{ 85.3$\pm$2.30                                    } & \textcolor{black}{ 82.75$\pm$3.41                                    } & \textcolor{black}{ 88.42$\pm$2.56    }                                \\\hline
\textit{IntraFusion}          & 10-fold           & 85.62$\pm$2.32                       & 85.62$\pm$2.32                      & 85.60$\pm$2.30                       & 85.70$\pm$2.23                       & 85.36$\pm$3.40                       & 10-fold              & 81.70$\pm$4.89 &
81.64$\pm$4.94 &
82.08$\pm$4.90 &
74.76$\pm$5.83 &
89.41$\pm$4.86                       \\
\textcolor{black}{\textit{InterFusion}   }       & 10-fold  &   84.87$\pm$2.23 &
84.85$\pm$2.25 &
84.90$\pm$2.34 &
80.71$\pm$3.16 &
85.09$\pm$2.10       & 10-fold & 84.67$\pm$2.67 &
84.62$\pm$2.70 &
84.93$\pm$2.97 &
77.80$\pm$3.29 &
90.07$\pm$4.04    \\
\textit{EF-LSTM}              & 10-fold           & 80.14$\pm$3.18                       & 79.95$\pm$3.20                      & 80.51$\pm$3.07                       & 89.81$\pm$6.12                       & 69.21$\pm$3.80                       & 10-fold              & 85.57$\pm$3.35 &
85.53$\pm$3.40 &
85.74$\pm$3.57 &
82.10$\pm$4.08 &
89.38$\pm$8.12\\
\textit{LF-LSTM}              & 10-fold           & 79.14$\pm$2.47                       & 79.05$\pm$2.59                      & 79.30$\pm$2.73                       & 85.77$\pm$4.25                       & 72.83$\pm$6.95                       & 10-fold              & 85.56$\pm$3.92 &
85.55$\pm$3.96 &
85.74$\pm$3.96 &
82.95$\pm$6.40 &
88.53$\pm$4.45                       \\ \hline
\textit{IntraAtt}             & 10-fold           & 87.70$\pm$0.92                       & 87.69$\pm$0.91                      & 87.89$\pm$1.01                       & 90.76$\pm$3.07                       & 84.02$\pm$1.40                       & 10-fold              & 86.60$\pm$4.61 &
86.57$\pm$4.63 &
86.84$\pm$4.56 &
83.17$\pm$7.65 &
89.51$\pm$4.97\\
\textit{InterAtt}             & 10-fold           & 87.38$\pm$1.80                       & 87.39$\pm$1.79                      & 87.42$\pm$1.72                       & 87.31$\pm$1.88                       & 87.54$\pm$3.08                       & 10-fold              & 86.47$\pm$4.06 &
86.45$\pm$4.08 &
86.54$\pm$3.95 &
83.01$\pm$6.29 &
90.08$\pm$2.00 \\
\textit{AuD-Former}           & 10-fold           & \textbf{91.13$\pm$1.93}              & \textbf{91.14$\pm$1.87}             & \textbf{91.16$\pm$1.95}              & \textbf{91.11$\pm$2.83}              & \textbf{91.22$\pm$1.80}              & 10-fold              & \textbf{88.53$\pm$1.09} &
\textbf{88.55$\pm$1.09} &
\textbf{88.68$\pm$1.13} &
\textbf{87.15$\pm$1.04} &
\textbf{90.22$\pm$3.14}
\\ \hline
Dataset                       & \multicolumn{6}{c||}{IPVS}                                                                                                                                                                                           & \multicolumn{6}{c}{\textcolor{black}{PC-GITA}}                                                                                                                                                                                            \\
Metric                        & \textit{EM} & \textit{ACC(\%$^h$)} & \textit{F1(\%$^h$)} & \textit{AUC(\%$^h$)} & \textit{SEN(\%$^h$)} & \textit{SPE(\%$^h$)} & \textit{EM}    & \textit{ACC(\%$^h$)} & \textit{F1(\%$^h$)} & \textit{AUC(\%$^h$)} & \textit{SEN(\%$^h$)} & \textit{SPE(\%$^h$)} \\ \hline
\textit{CNN-EMD$^*$}          & 5-fold            & 73.76                                & -                                   & -                                    & 73.14                                & 74.94                                & -                    & -                                    & -                                   & -                                    & -                                    & -                                    \\
\textit{Hybrid U-Lossian$^*$} & -                 & 89.64                                & 89.74                               & 94.33                                & \textbf{95.43}                       & 84.40                                & -                    & -                                    & -                                   & -                                    & -                                    & -                                    \\
\textit{NCA+SVM$^*$}                           & -                 & -                                    & -                                   & -                                    & -                                    & -                                    & -               & 82.03$\pm$2.67                                & -                                   & -                                    & -                                & -                                \\
\textit{QCP Glottal flow$^*$}                            & -                 & -                                    & -                                   & -                                    & -                                    & -                                    & 10-fold                    &  68.56$\pm$0.87                                &  -                               & -                                &  63.40$\pm$2.48                       &  73.73$\pm$3.01                                \\ \hline
\textcolor{black}{\textit{MM-Score$\vartriangle$} }        & 10-fold            & \textcolor{black}{ 75.25$\pm$4.75                                    } & \textcolor{black}{ 78.10$\pm$5.90                               } & \textcolor{black}{ 76.36$\pm$4.31                                } & \textcolor{black}{ 76.32$\pm$4.51                                } & \textcolor{black}{ 77.36$\pm$4.99                                    } & \textcolor{black}{ 10-fold                    } & \textcolor{black}{ 65.47$\pm$6.36                                    } & \textcolor{black}{ 63.62$\pm$6.16                                   } & \textcolor{black}{ 65.3$\pm$6.07                                    } & \textcolor{black}{ 57.25$\pm$5.90                                    } & \textcolor{black}{ 67.80$\pm$6.10}                                    \\
\textcolor{black}{\textit{FAIR$\vartriangle$}    }     & 10-fold            & \textcolor{black}{82.89$\pm$5.10                                    } & \textcolor{black}{ 84.32$\pm$5.82                                   } & \textcolor{black}{ 85.64$\pm$5.19                        } & \textcolor{black}{84.25$\pm$6.36                        } & \textcolor{black}{80.66$\pm$5.77                        } & \textcolor{black}{ 10-fold                    } & \textcolor{black}{ 66.53$\pm$8.20                                    } & \textcolor{black}{ 65.25$\pm$8.45                                   } & \textcolor{black}{ 66.28$\pm$7.88                                    } & \textcolor{black}{ 60.48$\pm$7.22                                    } & \textcolor{black}{ 68.18$\pm$6.77   }                                 \\\hline
\textit{IntraFusion}          & 10-fold           & 92.76$\pm$7.32                       & 93.06$\pm$6.78                      & 93.53$\pm$6.15                       & 92.92$\pm$9.82                       & 94.13$\pm$9.71                       & 10-fold             & 70.21$\pm$7.07 &
69.05$\pm$7.40 &
70.03$\pm$9.60 &
60.77$\pm$21.20 &
79.29$\pm$20.74\\
\textcolor{black}{\textit{InterFusion}  }        & 10-fold          & 90.08$\pm$2.53 &
89.72$\pm$6.29 &
86.17$\pm$3.5 &
82.01$\pm$4.24 &
80.32$\pm$9.36      & 10-fold & 72.11$\pm$7.48 &
72.04$\pm$6.73 &
73.05$\pm$7.93 &
72.19$\pm$13.75 &
73.90$\pm$14.79    \\
\textit{EF-LSTM}              & 10-fold           & 81.15$\pm$5.36                       & 78.97$\pm$9.56                      & 75.77$\pm$13.75                      & 66.40$\pm$21.25                      & 93.14$\pm$10.88                      & 10-fold       & 60.04$\pm$18.97 &
57.21$\pm$14.49 &
55.24$\pm$10.48 &
57.14$\pm$46.95 &
53.33$\pm$45.22            \\
\textit{LF-LSTM}              & 10-fold           & 75.75$\pm$7.40                       & 76.38$\pm$5.71                      & 77.26$\pm$4.60                       & 77.73$\pm$21.24                      & 76.80$\pm$19.68                      & 10-fold              & 63.33$\pm$12.47 &
62.39$\pm$10.67 &
59.35$\pm$8.98 &
80.48$\pm$13.47 &
32.22$\pm$25.43    \\ \hline
\textit{IntraAtt}             & 10-fold           & 92.35$\pm$3.45                       & 92.60$\pm$3.12                      & 93.35$\pm$2.20                       & 92.70$\pm$5.36                       & 94.00$\pm$5.75                       & 10-fold           &   79.33$\pm$4.71 &
78.57$\pm$5.30 &
79.56$\pm$8.03 &
80.24$\pm$4.38 &
58.89$\pm$20.43         \\
\textit{InterAtt}             & 10-fold           & 94.70$\pm$6.17                       & 94.31$\pm$6.93                      & 92.20$\pm$10.59                      & 86.61$\pm$9.76                       & 96.48$\pm$7.44                       & 10-fold              & 78.31$\pm$7.70 &
73.13$\pm$4.22 &
78.21$\pm$8.75 &
70.71$\pm$15.02 &
75.71$\pm$12.95     \\
\textit{AuD-Former}           & 10-fold           & \textbf{96.39$\pm$1.60}              & \textbf{96.44$\pm$1.54}             & \textbf{95.84$\pm$2.60}              & 94.20$\pm$6.52              & \textbf{97.78$\pm$3.09}              & 10-fold              & \textbf{84.67$\pm$4.71} & \textbf{83.94$\pm$4.32} & \textbf{84.13$\pm$2.51} & \textbf{82.81$\pm$9.91} &\textbf{87.33$\pm$7.86 }         \\ \hline
\end{tabular}}
\end{table*}

\begin{table}[]
\centering
\caption{\textcolor{black}{Summary of experimental results on the SVD dataset in terms of average and standard deviations of Accuracy (ACC), F1 score, Area Under Curve (AUC), Sensitivity (SEN), and Specificity (SPE). \textit{EM}: Evaluation Method; $^h$: higher means better; $^*$: reported from literature; $\vartriangle$: re-implemented; -: not reported. } }
\label{table:SVD}
 \resizebox{\linewidth}{!}{
\begin{tabular}{c||cccccc}
\hline
\multirow{2}{*}{\begin{tabular}[c]{@{}c@{}}dataset\\Metric\end{tabular}} & \multicolumn{6}{c}{SVD}                                                                                                                                                                                                                                                                                                                                                                                                                                                                  \\
                                                                             & \textit{EM} & \textit{ACC(\%$^h$)} & 
                                                                        \textit{F1(\%$^h$)}  &   \textit{AUC(\%$^h$)} & \textit{SEN(\%$^h$)} & \textit{SPE(\%$^h$)}  \\ \hline
\textit{DW+CLL+CNL$^*$}                                         & 10-fold                      & 70.77$\pm$1.05                                                                         & -                                                                                     & -                                                                                      & -                                                                                      & -                                                                                      \\
\textit{Resnet18+SVM$^*$}                                  & -                            & 80.9                                                                                   & -                                                                                     & -                                                                                      & -                                                                                      & -                                                                                      \\ \hline
\textcolor{black}{\textit{MM-Score$\vartriangle$}}                                        & 10-fold & \textcolor{black}{73.20$\pm$4.35 } & \textcolor{black}{
72.06$\pm$3.82 } & \textcolor{black}{
73.69$\pm$3.70 } & \textcolor{black}{
73.89$\pm$5.31 } & \textcolor{black}{
72.17$\pm$5.41 }\\
\textcolor{black}{\textit{FAIR$\vartriangle$}   }                                     & 10-fold & \textcolor{black}{75.79$\pm$3.68 } & \textcolor{black}{
75.75$\pm$2.92 } & \textcolor{black}{
78.62$\pm$3.96 } & \textcolor{black}{
79.27$\pm$2.58 } & \textcolor{black}{
75.72$\pm$3.29}\\\hline
\textit{IntraFusion}                                        & 10-fold & 73.59$\pm$3.44 &
73.58$\pm$3.43 &
73.63$\pm$3.53 &
72.99$\pm$5.30 &
74.26$\pm$4.75\\
\textit{InterFusion}                                        & 10-fold & 76.06$\pm$1.91 &
76.03$\pm$1.93 &
76.19$\pm$1.80 &
79.60$\pm$2.35 &
72.79$\pm$5.10\\
\textit{EF-LSTM}                                   & 10-fold &                              72.49$\pm$3.79 &
71.85$\pm$4.36 &
70.75$\pm$4.73 &
62.49$\pm$16.21 &
79.02$\pm$10.75 \\
\textit{LF-LSTM}                                            &   10-fold                           & 74.19$\pm$2.90 &
72.16$\pm$4.08 &
69.26$\pm$4.95 &
58.68$\pm$13.12 &
79.89$\pm$5.14\\ \hline
\textit{IntraAtt}                                           & 10-fold & 78.39$\pm$2.38 &
78.01$\pm$2.16 &
76.03$\pm$2.19 &
64.81$\pm$4.76 &
87.26$\pm$5.51\\
\textit{InterAtt}                                           &    10-fold &
77.98$\pm$5.50 &
76.87$\pm$6.26 &
75.29$\pm$6.98 &
62.44$\pm$17.86 &
\textbf{88.14$\pm$9.77}\\
\textit{AuD-Former}                                         &    10-fold                          & \textbf{82.27$\pm$2.29} &
\textbf{82.21$\pm$2.27} &
\textbf{82.35$\pm$2.01} &
\textbf{84.68$\pm$4.89} &
80.03$\pm$3.75\\ \hline                                                                                   
\end{tabular}}
\end{table}

\section{Results and Analysis}

\subsection{Quantitative Measurements}
\subsubsection{Comparisons against Baselines} 
Tables~\ref{table:baselines} and \ref{table:SVD} present the results of the performance comparison between our \textit{AuD-Former} and other state-of-the-art multimodal fusion baselines on the all datasets. It can be observed that our \textit{AuD-Former} surpasses all baselines across all metrics during cross-validation experiments \textcolor{black}{in diagnosing diverse diseases, including COVID-19, PD, and pathological dysarthria. This indicates that, in terms of overall performance, the \textit{AuD-Former} has more promising potential as a robust benchmark model for general audio-based disease detection tasks.
Further details of the comparison to answer the proposed research question are summarized as follows:}

\noindent \textcolor{black}{\textit{\textbf{Effectiveness of the Hierarchical Fusion Strategy.}}} The experimental results presented in Tables~\ref{table:baselines} and \ref{table:SVD} demonstrate that our \textit{AuD-Former} significantly outperforms all reported baselines with unilateral fusion strategies. Specifically, when compared to baselines that utilize only intra-modal fusion—such as \textcolor{black}{\textit{AE+RF}} on the Coswara dataset, \textit{CNN-EMD}, \textit{Hybrid U-Lossian} on the IPVS, \textit{NCA+SVM} on the PC-GITA, and \textit{DW+CLL+CNL}, \textit{Resnet18+SVM} on the SVD—\textit{AuD-Former} shows an absolute improvement across all evaluation metrics by $2.64\% - 44.11\%$. While the \textit{Hybrid U-Lossian} baseline exhibits a higher performance in terms of sensitivity, this outperformance is not substantiated through subject-independent cross-validation, and leads to a significant lower specificity. Such performance improvements of the \textit{AuD-Former} are reasonable since these intra-modal fusion baselines neglect the complementary information that can benefited from the fusion across different bio-acoustic modalities. 
 \begin{figure*}[t]
    \centering
    \includegraphics[width=0.95\linewidth]{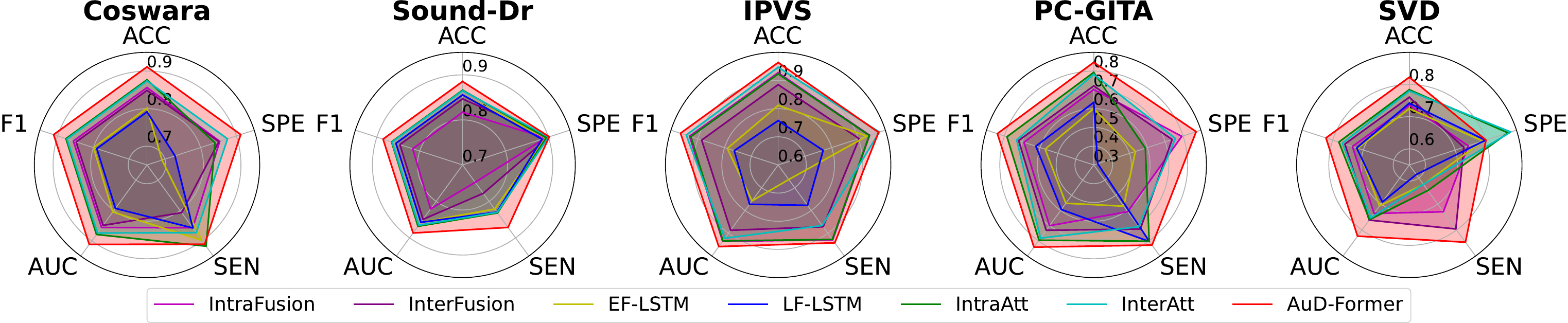}
    \caption{\textcolor{black}{Comparative visualization of the \textit{AuD-Former} with implemented baselines (\textit{IntraFusion}, \textit{IntraFusion}, \textit{EF-LSTM}, and \textit{LF-LSTM}) and ablation models (\textit{InterAtt} and \textit{InterAtt}).} }
    \label{fig:RESULT}
    \vspace{-15pt}
    \end{figure*}

Similarly, \textit{AuD-Former} significantly outperforms baselines that utilize only inter-modal fusion—such as \textit{MM-Score} for all datasets and \textit{QCP Glottal flow} on the PI-GITA. This advantage is largely because these baselines neglect the benefits of intra-modal fusion, which is essential for capturing detailed and nuanced correlations within each bio-acoustic modality before their cross-modal integration.

\textcolor{black}{More importantly, as shown in Fig. \ref{fig:RESULT}, Tables~\ref{table:baselines} and \ref{table:SVD}, when compared to the \textit{IntraFusion} and \textit{InterFusion} baselines, which retain the intra-modal representation learning module and the full hierarchical transformer network, respectively, \textit{AuD-Former} still consistently excels across all performance indicators on all datasets. This superiority can be attributed to the fact that \textit{IntraFusion} solely relies on intra-modal fusion within a single bio-acoustic modality, and \textit{InterFusion} limits itself to integrating features across modalities without adequately considering intra-modal interactions.} 

Overall, these observations underscore the critical limitations of unilateral strategies: they fail to simultaneously harness the complementary nature of different feature domains within and across bio-acoustic modalities, even when extensive independent exploration of intra- or inter-modal dependencies is conducted. Our approach demonstrates its superiority not only through advanced attention mechanisms but also by implementing a more comprehensive strategy for feature and modality integration. \textcolor{black}{By systematically integrating intra-modal and inter-modal fusion, \textit{AuD-Former} effectively captures and combines complementary and relevant information from multiple modalities. This enables the model to generate a unified multimodal representation that improves the accuracy and robustness of disease prediction.}

\noindent \textit{\textbf{Effectiveness of the Intra-modal and Inter-modal Representation Learning.}} Fig. \ref{fig:RESULT}, Tables~\ref{table:baselines} and \ref{table:SVD} demonstrate that \textit{AuD-Former} consistently outperforms baselines such as \textit{FAIR}, \textit{EF-LSTM}, and \textit{LF-LSTM} across all datasets. Despite these baselines also employ both intra-modal and inter-modal fusion strategies, their primary limitation lies in their simplistic approach to learn unimodal or multimodal representations during these fusions. Specifically, these baseline models typically rely on straightforward feature concatenation for intra-modal fusion. For inter-modal fusion, they either concatenate features from each modality before processing them through a self-attention or LSTM layer or concatenate the outputs post-LSTM processing. \textcolor{black}{In contrast, \textit{AuD-Former} utilizes a hierarchical Transformer structure that considers both intra-modal and inter-modal dependencies during the hierarchical fusion phases.} This sophisticated approach enables more effective extraction of comprehensive unimodal and multimodal representations, significantly enhancing disease prediction performance.

\textcolor{black}{Moreover, we can observe that even the \textit{IntraFusion} and \textit{InterFusion} baselines, which only focus on a single level of fusion, show superior performance over \textit{EF-LSTM} and \textit{LF-LSTM} across most datasets, further underscoring the limitations of simpler representation learning. This observation confirms that while hierarchical fusion strategies introduce potential for improved performance, inadequate exploration of dependencies within modality-specific and modality-shared spaces can lead to inefficient unimodal and multimodal representations, thereby limited performance enhancements.}

\textcolor{black}{Overall, these results highlight the benefits of intra-modal and inter-modal representation learning in \textit{AuD-Former} in addition to the hierarchical fusion strategy. The proposed hierarchical Transformer based layers can effectively capture the intricate dependencies within and across modalities, enabling the model to learn more expressive and informative unimodal and multimodal representations. }

\subsubsection{Results of the Ablation Study}
\label{ABS}
As illustrated in Table~\ref{table:baselines} and Fig. \ref{fig:RESULT}, our \textit{AuD-Former} outperforms ablation models \textit{IntraAtt} and \textit{InterAtt} by $1.35\% - 9.20\%$ on Coswara dataset, by $0.14\% - 4.14\%$ on Sound-Dr dataset, by $1.00\% - 7.59\%$ on IPVS dataset, by $2.57\% - 28.44\%$ on PC-GITA dataset, and by $3.88\% - 22.24\%$ across most metrics averagely. While the \textit{IntraAtt} demonstrates notable sensitivity on the Coswara dataset, the \textit{InterAtt} model shows respectable specificity on the IPVS dataset, and both of them exhibit higher specificity on the SVD dataset, these successes are offset by significant drawbacks in their respective counterbalancing metrics. Such disproportionate performance of these two ablation models could compromise decision-making procedures. For instance, within a healthcare context, a surge of false positives from the \textit{IntraAtt} model might prompt unnecessary treatments, while an increase in false negatives from the \textit{InterAtt} model might result in overlooked diagnoses.

In contrast, our \textit{AuD-Former} model manages to maintain a balanced and superior performance across these metrics, resulting in an overall enhanced performance in terms of ACC, AUC, and F1. This improvement can be rationalized by the fact that, although both ablation models adopt the same hierarchical fusion pattern as the \textit{AuD-Former}, they fail to fully exploit complementary dependencies in both modality-specific and modality-shared spaces due to the omission of either intra-modal or inter-modal representation learning modules. 

This observation underscores the simultaneous utilization of our proposed intra-modal and inter-modal attention modules, which contribute significantly to the superior performance of the \textit{AuD-Former}. \textcolor{black}{By effectively modeling complementary dependencies both within and across modalities, \textit{AuD-Former} generates a fusion representation that captures relevant multimodal features, improving its performance in audio-based disease prediction tasks.}

    \begin{figure*}[t]
    \centering
    \includegraphics[width=\linewidth]{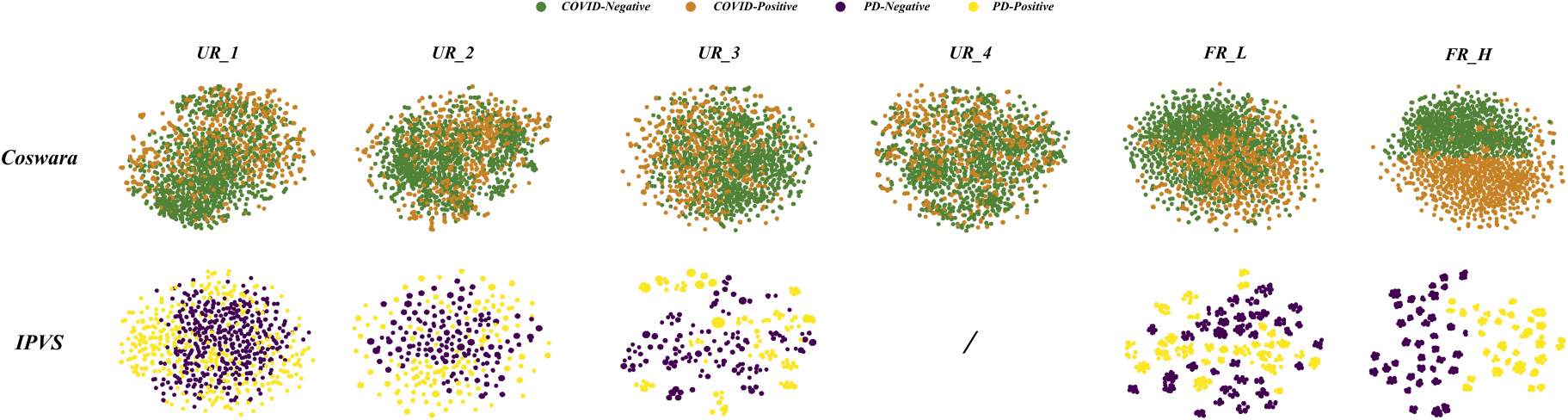}
    \caption{A t-SNE visualization of the learned representations within each modality-specific space, denoted as $\textcolor{black}{\textit{UR}}_{m}$, as well as the low-level and high-level modality-shared spaces, represented as $\textcolor{black}{\textit{FR}}_{L}$ and $\textcolor{black}{\textit{FR}}_{H}$, in the \textit{AuD-Former} respectively. }
    \label{fig:TSNE}
    \vspace{-10pt}
    \end{figure*}
    \begin{figure}[t]
    \centering
    \includegraphics[width=\linewidth]{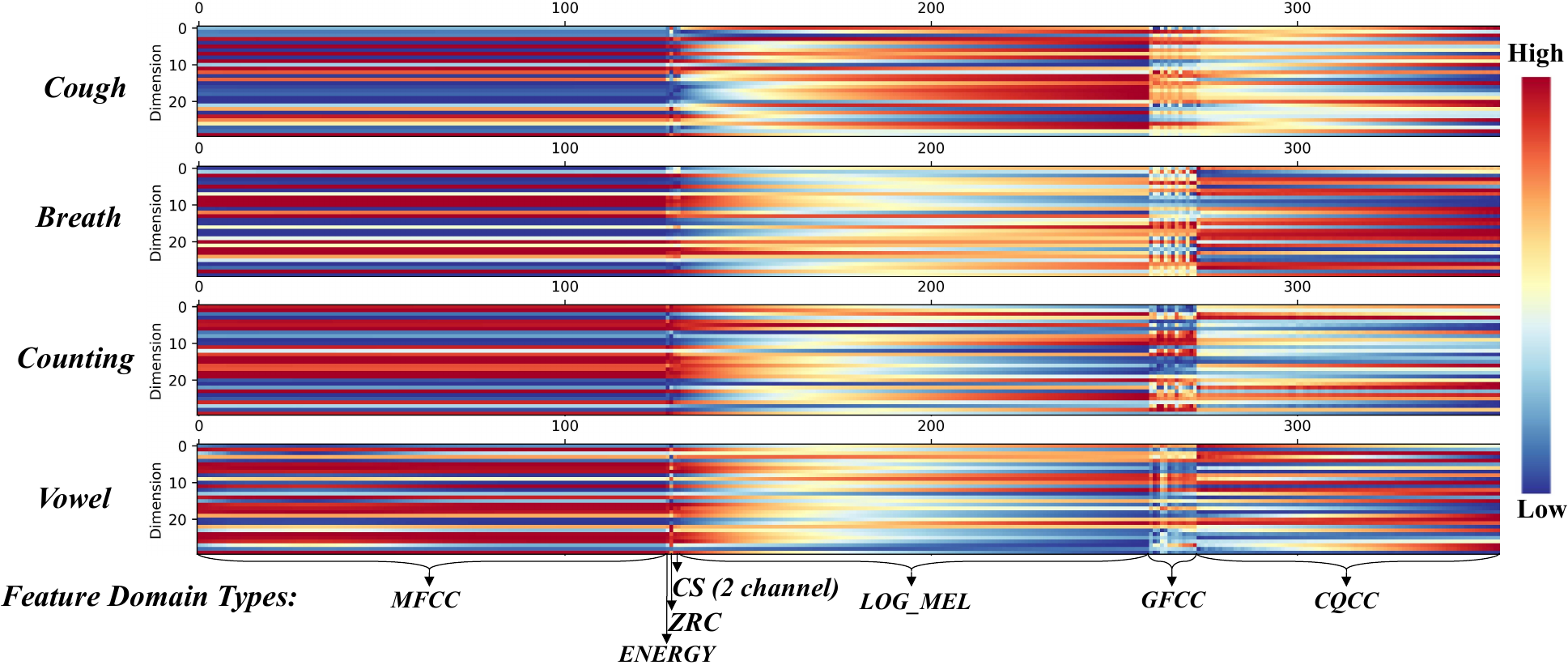}
    \caption{\textcolor{black}{Visualization of unimodal representations ${\textit{UR}}_m$ weighted by the learned intra-modal attention transformers on the Coswara dataset. Each block corresponds to a minimum unit generated by the temporal and positional embedding layers, representing different features within the learned unimodal representation. The dimensions of the visualized representations are $l_m \times d$, which is $356 \times 30$ for the Coswara dataset. Channel unit information is indicated along the x-axis.}}
    \label{fig:self-v}
    \vspace{-15pt}
    \end{figure}

    \begin{figure}[t]
    \centering
    \includegraphics[width=0.95\linewidth]{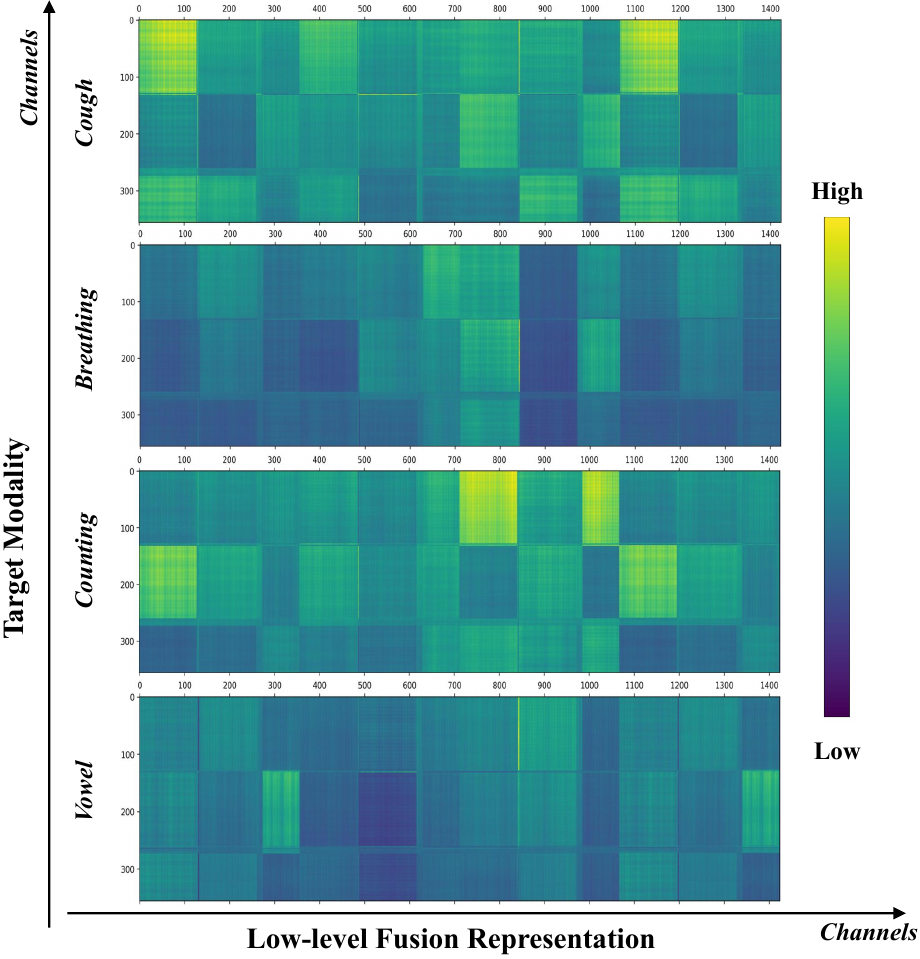}
    \caption{\textcolor{black}{Visualization of the cross-modal correlations learned between each target unimodal representation $\textit{UR}_m$ and the fused one ${\textit{FR}}_L$ in the Coswara dataset. Each block represents a cross-modal attention score learned between the low-level fusion representation and the target modality. The dimensions of each resultant cross-attention matrix are $l_m \times l_f$, corresponding to $356 \times 1424$ for the Coswara dataset. Along the x-axis, for $l_m$: the first $356$ units represent channels for unimodal representations of the cough modality, followed by $365$ units each for breathing, counting, and vowel modalities.} }
    \label{fig:cross-v}
    \vspace{-15pt}
    \end{figure}

\subsection{Qualitative Analysis}

In general, a proficient representation learning approach should facilitate a reliable and efficient encapsulation of the original patient data within the devised representation spaces. To provide an intuitive illustration of the effectiveness of the hierarchical representations learned in the \textit{AuD-Former} (as depicted in Fig. \ref{fig:framework}), namely the unimodal representations $\textcolor{black}{\textit{UR}}_m$ in the modality-specific spaces, low-level fusion representation $\textcolor{black}{\textit{FR}}_L$, and high-level fusion representation $\textcolor{black}{\textit{FR}}_H$ in the modality-shared spaces, we mapped these representations into a two-dimensional space using the t-SNE method \cite{van2008visualizing}. As depicted in Fig. \ref{fig:TSNE}, we can observe that clusters representing two classes: healthy vs. ill, on each dataset become increasingly distinctive when moving from modality-specific representation spaces to the high-level modality-shared representation space. This proves that with the hierarchical structure implemented in our \textit{AuD-Former} to explore intra-modal and inter-modal correlations, a powerful multimodal representation used for downstream disease prediction tasks can be effectively learned. Additionally, it is evident that the high-level fusion representation is more effective than the low-level one, which is directly concatenated from unimodal representations. This observation further validates our proposed inter-modal fusion strategy, demonstrating its ability to produce superior, integrated representations. It underscores the value of employing sophisticated fusion strategies, such as the one implemented in our \textit{AuD-Former}, to ensure more robust and discriminative representations for improved performance in audio-based disease prediction tasks.

To further demonstrate how the intra-modal representation learning module works, we visualized the unimodal representation for each modality, as produced by each intra-modal transformer on the Coswara dataset. Each block of the visualized unimodal representations in Fig. \ref{fig:self-v} represents a minimum unit (generated by the temporal and positional embedding layers) of different unimodal features. The values in each unit are normalized through the feature dimension using Min-Max normalization for better visualization. \textcolor{black}{We can observe distinct color distributions and clear color stratification across different feature domains within the same modality, demonstrating that the intra-modal attention has learned to differentiate and assign unique attention weights to each feature domain when generating the unified unimodal representation. Moreover, within each feature domain, we observe uniform color intensities along the temporal dimension (x-axis), suggesting that the intra-modal module has learned to maintain consistent temporal dependencies during feature processing rather than treating each time step independently.}

\begin{figure*}[t]
\centering
\includegraphics[width=\linewidth]{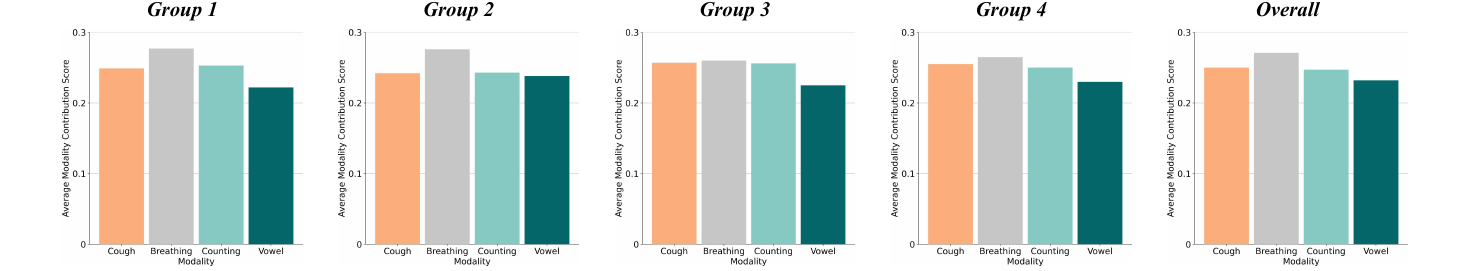}
\caption{Case visualizations of the average modality contribution score learned by the \textit{AuD-Former} in the final decision representation on the Coswara dataset. \textcolor{black}{Groups 1-4 represent random subsets of 10 patients each, while the Overall shows the average across all patients in the dataset.}}
\vspace{-15pt}
\label{fig:MCS}
\end{figure*}

Moreover, to demonstrate the effectiveness of the inter-modal representation learning module, we visualized the cross-modal attention matrix between each modality and others, as learned by each cross-modal transformer, on the Coswara dataset. \textcolor{black}{As depicted in Fig. \ref{fig:cross-v}, we observe that each modality does not necessarily exhibit the highest correlation score with itself. This is because the cross-modal attention mechanism encourages modalities to leverage complementary information from others, while the prior intra-modal attention layers have made each modality more self-sufficient. Instead, the cross-modal attention patterns vary across modalities, emphasizing the model's ability to prioritize highly complementary information (represented by yellow) and downplay irrelevant information (depicted in purple) at various positions in the low-level fusion representations. These patterns illustrate the dynamic behavior of the cross-attention mechanism, which enables the model to effectively combine relevant information across modalities and enhance its overall performance in capturing inter-modal relationships.}

Additionally, to comprehend the decision-making process of our model, we delved into the final contribution of each modality as learned by \textit{AuD-Former}, taking the Coswara dataset as a case study. For this purpose, we visualized the modality contribution score of each modality in the final fusion representation, which is generated by the last multi-head self-attention layer in the prediction layer (as discussed in Section \ref{Model}). Specifically, the modality contribution score (MCS) of the $i^{th}$ modality in the final representation of a patient $p$ can be computed as follows: $\operatorname{MCS_i^p}={SA_{i}}/{\sum_{m=1}^{M}{SA_{m}}}$, where $M$ is the number of total modalities, $SA_{i}$ refers to the attention score assigned to the $i^{th}$ modality during the multi-head self-attention process in the prediction layer, \textcolor{black}{calculated as: $SA_i = \frac{1}{H} \sum_{h=1}^H \frac{1}{l_i} \sum_{j=s_i}^{e_i} \sum_{k=1}^{N} \text{softmax}\left(\frac{Q_j^h \cdot (K_k^h)^\top}{\sqrt{d_k}}\right)$, where $H$ is the number of attention heads, $l_i$ is the length of the $i$-th modality's representation, $s_i$ and $e_i$ are its start and end indices in the multimodal representation $FR_H$, $N$ is the total sequence length, and $Q_j^h$ and $K_k^h$ are rows of the query and key matrices for the $h$-th head, respectively.}

\textcolor{black}{We sampled four separate groups from the Coswara dataset, with each group consisting of ten randomly selected patients. For each group, we calculated the average MCS per patient. Additionally, we computed the average MCS of all patients in the dataset for comparison.} As shown in Fig. \ref{fig:MCS}, the average MCS across the four groups closely mirrors the overall score distribution, revealing some consistent patterns. For instance, the breathing modality persistently receives a higher level of importance in the decision-making process for various patients. This suggests that more valuable information can be extracted for diagnosing COVID-19, aligning with findings from previous studies that breathing signals can better reflect the state of the lungs and the pulmonary vasculature \cite{dhawan2021beyond}. Similarly, the vowel modality consistently receives less importance, echoing its inconsistent performance when using the \textit{IntraFusion} model, as shown in Appendix \ref{appendix:B}. Notably, these patterns demonstrate some degree of specificity for each group of patients. This suggests that our \textit{AuD-Former} is capable of learning a decision representation that is both general and patient-specific, thus providing an interpretable basis for diagnostic decisions for each patient.

\subsection{\textcolor{black}{Discussion}} \textcolor{black}{Based on the observations discussed above, the research question posed earlier can be affirmatively answered. The hierarchical integration of fusion strategies, encompassing both the fusion of different feature domains within a single modality and the fusion across various modalities, can significantly enhance the performance of audio-based disease prediction tasks by learning a more informative multimodal representation. This enhancement is evidenced by the superior results achieved by \textit{AuD-Former} compared to baselines that employ unilateral fusion strategies, which focus solely on either intra-modal or inter-modal fusion.}

\textcolor{black}{However, it is crucial to emphasize that the performance enhancement is contingent upon the simultaneous and comprehensive exploration of latent intra-modal and inter-modal dependencies. Insufficient or isolated exploration of these dependencies may lead to suboptimal unimodal and multimodal representations, which can hinder predictive performance when employing hierarchical fusion. This is evidenced by the performance decline observed in baselines or ablation models with independent or inadequate exploration of intra- and inter-modal dependencies, while implementing the hierarchical fusion strategy with the same multimodal feature inputs as \textit{AuD-Former}. In contrast, as shown in Fig. \ref{fig:TSNE}, the hierarchical representation learning modules in \textit{AuD-Former} can learn increasingly effective representations as the hierarchical fusion progresses, from unimodal features to unimodal representations and finally to comprehensive multimodal representations. }

\textcolor{black}{To sum up, the experimental results confirm that the hierarchical integration of intra-modal and inter-modal fusion processes, along with the concurrent and thorough exploration of latent dependencies within both modality-specific and modality-shared spaces, can effectively query informative multimodal representations using unimodal feature sets. Consequently, the \textit{AuD-Former} framework stands out as a promising approach for leveraging the complementary nature of different feature domains and modalities, setting the stage for more accurate and robust audio-based disease prediction systems.}

\section{Conclusion and Future Work}
In this work, we present \textit{AuD-Former}, a hierarchical transformer network for multimodal audio-based disease prediction. By hierarchically leveraging intra-modal and inter-modal fusion strategies, \textit{AuD-Former} captures dependencies within and across modalities, creating a unified representation for disease prediction without extensive feature selection. Experiments on three diseases (COVID-19, pathological dysarthria, and Parkinson’s) demonstrate the effectiveness of this approach.

Despite the promising results, translating these findings to real clinical settings remains a challenge. Future work will focus on optimizing \textit{AuD-Former} for real-world applications, improving adaptability to diverse patient demographics, and exploring its utility for predicting conditions such as mild cognitive impairment or early dementia. Additionally, we aim to investigate its potential for non-medical tasks, such as audio event detection and multimedia content analysis, which could benefit from similar hierarchical multimodal fusion strategies.

\bibliography{main}
\bibliographystyle{IEEEtran}

\appendices
\section{Experimental details of the \textit{AuD-Former}} \label{appendix:A}

\begin{table}[h]
    \centering
    \caption{Hyperparameters of the \textit{AuD-Former} and ablation models used in the experiments conducted on the Coswara and IPVS datasets.}
    \label{hypara}
     \resizebox{\linewidth}{!}{
    \begin{tabular}{c||c||c||c||c||c}
        \hline
        Parameter name & Coswara & IPVS &Sound-Dr &PC-GITA &SVD\\ \hline\hline
        Batch Size                    & 32    & 16     &16    &16   &16  \\
        Initial Learning Rate         & 1e-3  & 1e-3   &1e-3  &1e-4 &1e-3\\
        Optimizer                     & SGD   & SGD    &SGD   &SGD &SGD    \\
        Transformer Hidden Unit Size  & 40    & 40     &40    &40   &40\\
        Crossmodal Attention Heads    & 5     & 5      & 5    & 3   & 3      \\
        Crossmodal Attention Block Dropout & 0.1   & 0.1  & 0.1   & 0.1  & 0.1 \\
        Output Dropout                & 0.1    & 0.1   & 0.1  & 0.1 & 0.1\\
        Epochs                        & 60    & 100    &60    &80   &80 \\ \hline
    \end{tabular}}
\end{table}

\section{Peformance of each modality in the Coswara and IPVS datasets with intra-modal fusion models} \label{appendix:B}

\begin{table}[htb]
\centering
\caption{Peformance of each modality with the best-performing intramodal fusion model. $^\clubsuit$: classification with Transformer; $^\spadesuit$: classification with GAT.}
\label{table:single modality}
\resizebox{0.5\textwidth}{!}{
\begin{tabular}{c||ccccc}
\hline
Dataset           & \multicolumn{5}{c}{Coswara}                                                                                                                           \\ 
Metric            & \textit{ACC(\%$^h$)}  & \textit{F1(\%$^h$)} & \textit{AUC(\%$^h$)}  & \textit{SEN(\%$^h$)}    & \textit{SPE(\%$^h$)}  \\ \hline\hline
Cough        & 83.40$\pm$1.78$^\clubsuit$    & 83.37$\pm$1.79$^\clubsuit$    & 83.33$\pm$1.82$^\clubsuit$   & 81.49$\pm$3.88$^\clubsuit$    & 84.86$\pm$2.76$^\spadesuit$               \\
Breath       & 85.62$\pm$0.54$^\spadesuit$   & 85.60$\pm$0.51$^\spadesuit$   & 85.70$\pm$0.58$^\spadesuit$  & 82.25$\pm$2.68$^\spadesuit$   & 86.44$\pm$4.53$^\spadesuit$               \\
Counting     & 84.62$\pm$2.32$^\spadesuit$   & 84.62$\pm$2.32$^\spadesuit$   & 84.60$\pm$2.30$^\spadesuit$  & 84.70$\pm$2.23$^\spadesuit$   & 84.36$\pm$3.40$^\spadesuit$               \\
Vowel        & 79.93$\pm$2.43$^\clubsuit$    & 79.90$\pm$2.47$^\clubsuit$    & 79.96$\pm$2.44$^\clubsuit$   & 77.63$\pm$5.93$^\clubsuit$    & 79.18$\pm$5.59$^\clubsuit$                \\ \hline\hline
Dataset           & \multicolumn{5}{c}{IPVS}                                                                                                                              \\ 
Metric            & \textit{ACC(\%$^h$)}  & \textit{F1(\%$^h$)} & \textit{AUC(\%$^h$)}  & \textit{SEN(\%$^h$)}    & \textit{SPE(\%$^h$)}  \\ \hline\hline
Text Reading        & 83.03$\pm$22.43$^\clubsuit$   & 79.72$\pm$29.29$^\clubsuit$   & 85.76$\pm$17.98$^\clubsuit$   & 77.35$\pm$38.72$^\clubsuit$   & 94.17$\pm$3.80$^\clubsuit$               \\
Phrase Reading       & 79.38$\pm$13.69$^\spadesuit$  & 79.17$\pm$13.43$^\spadesuit$  & 79.16$\pm$20.69$^\clubsuit$   & 69.26$\pm$36.91$^\clubsuit$   & 89.05$\pm$14.07$^\clubsuit$              \\
Syllable Pronunciation     & 92.76$\pm$7.32$^\clubsuit$    & 93.06$\pm$6.78$^\clubsuit$    & 93.53$\pm$6.15$^\clubsuit$    & 92.92$\pm$9.82$^\clubsuit$    & 94.13$\pm$9.71$^\clubsuit$               \\
\hline
\end{tabular}}
\end{table}

\end{document}